\documentclass[aip,jap,reprint]{revtex4-1}

\draft 

\usepackage{graphicx}
\usepackage{amsmath}
\graphicspath{{./FIGURES/}}

\begin{document}


\title{Semianalytical quantum model for graphene field-effect transistors} 



\author{Claudio Pugnaghi}
\author{Roberto Grassi}
\email[]{roberto.grassi@unibo.it}
\author{Antonio Gnudi}
\author{Valerio Di Lecce}
\author{Elena Gnani}
\author{Susanna Reggiani}
\author{Giorgio Baccarani}
\affiliation{ARCES and DEI, University of Bologna, Viale Risorgimento 2, 40136 Bologna, Italy}


\date{\today}

\begin{abstract}
We develop a semianalytical model for monolayer graphene field-effect transistors in the ballistic limit. Two types of devices are considered: in the first device, the source and drain regions are doped by charge transfer with Schottky contacts, while, in the second device, the source and drain regions are doped electrostatically by a back gate. The model captures two important effects that influence the operation of both devices: (\emph{i}) the finite density of states in the source and drain regions, which limits the number of states available for transport and can be responsible for negative output differential resistance effects, and (\emph{ii}) quantum tunneling across the potential steps at the source-channel and drain-channel interfaces. By comparison with a self-consistent non-equilibrium Green's function solver, we show that our model provides very accurate results for both types of devices, in the bias region of quasi-saturation as well as in that of negative differential resistance.
\end{abstract}


\onecolumngrid
Copyright (2014) American Institute of Physics. This article may be downloaded for personal use only. Any other use requires prior permission of the author and the American Institute of Physics. The following article appeared in J. Appl. Phys. \textbf{116}, 114505 (2014) and may be found at \url{http://dx.doi.org/10.1063/1.4895993}.
\vskip0.5cm
\twocolumngrid

\maketitle 

\section{\label{sec_intro}Introduction}

Thanks to its exceptional properties, graphene \cite{Geim07,CastroNeto09,Fuhrer10} has attracted the interest of the scientific community over the last decade, inaugurating a new line of research focusing on two-dimensional crystals for electronic and optoelectronic applications. Atomically thin 2-D materials open the way to new device concepts \cite{Britnell12,Vaziri13}, but they also promise to eliminate the short-channel effects that afflict conventional MOSFETs at the scaling limit \cite{Schwierz10}. Unfortunately, use of graphene field-effect transistors (GFETs) as a replacement for conventional semiconductor switches in digital circuits is made difficult by the zero-bandgap nature of the material, which causes Klein tunneling (band-to-band tunneling from the conduction to the valence band and vice-versa, with unusual tunneling probability equal to one at normal incidence \cite{Katsnelson06}) and prevents devices from being switched off. The situation is different for analog circuits, where the lack of a band gap is less critical and the very high carrier mobility of graphene at room temperature (up to $2\times10^5$ cm$^2$~V$^{-1}$~s$^{-1}$ for suspended graphene \cite{Bolotin2008}) and its high saturation velocity ($\sim4\times10^7$ cm/s, larger than the peak velocities of common III--V semiconductors \cite{Schwierz10}) are ideally suited for high-frequency operation. Evolution of GFETs targeting analog applications has been fast and devices with good performance in terms of cut-off frequency (comparable or higher than their CMOS counterparts) have been demonstrated \cite{Liao10,Wu11,Lin11}.

On the modeling side, the work done on GFETs has also been considerable. In the literature, one can find several compact or semianalytical models that can be calibrated to accurately reproduce the terminal characteristics of experimental devices \cite{Thiele10,Jimenez11,Rodriguez14}. Such models, being based on drift-diffusion equations, assume the diffusive limit and are therefore expected to provide a good picture of the device physics only for relatively long channels, where scattering is significant. Simulations of ultimately scaled GFETs have been performed in most cases using more sophisticated models, which combine quantum transport within the non-equilibrium Green's function (NEGF) formalism \cite{DattaQT2005}, using either an atomistic tight-binding \cite{Zhao11} or a Dirac Hamiltonian \cite{Chauhan12}, and self-consistent electrostatics. Such models are suitable for simulating the ballistic limit and, compared to semiclassical Monte-Carlo approaches \cite{Paussa14}, provide a rigorous treatment of Klein tunneling. On the other hand, they are computationally demanding. Simpler models that capture the essential physics of short-channel GFETs would be more handy for repeated use in device optimization studies where simulation speed, besides accuracy, is important.

Few semianalytical ballistic models for short-channel GFETs have actually been proposed in the past \cite{Koswatta11,Grassi13}. However, they are not completely satisfactory when compared to NEGF simulations. We note that, due to the ambipolar nature of transport in graphene, two transport regimes are possible in graphene devices: the regime of quasi-saturation (the characteristics do not saturate with increasing drain voltage $V_{D}$, but rather show an inflection point where the output differential conductance has a minimum) and the one of negative output differential resistance (NDR). Both are observed in experiments with long-channel devices \cite{Meric08,Wu12}, but are also possible in short-channel ones. The model in Ref.~\citenum{Koswatta11}, based on semiclassical transport, takes into account the peculiar electronic structure of graphene in the channel, but ignores possible variations in the electric potential (and thus, at a given energy, in the density of states (DOS)) between the channel and the source and drain regions. As a consequence, it describes well the quasi-saturation phenomenon but not NDR. The model presented in Ref.~\citenum{Grassi13} is based on semiclassical transport too, but includes the effect of the difference in potential/DOS between the channel and the source and drain regions. In particular, it accounts for the fact that transmission at a given energy is limited by the region where the DOS is minimum (``mode bottleneck effect''). The model has been tested with reference to devices with self-aligned contacts. Although it can capture the NDR effect, the agreement with NEGF is only qualitative. The reason lies in the simplified way in which transport across the potential steps at the source-channel and drain-channel interfaces is treated. In particular, Klein tunneling is included with tunneling probability equal to one, ignoring the quantum-mechanical effect of wavefunction mismatch at the junctions and also the fact that, if the junctions are not perfectly abrupt, electrons incident at non-normal angles need to tunnel through an apparent band gap \cite{Low12}. The latter effect should be more evident in devices with spacings between the gated part of the channel and the source and drain contacts (``gate underlaps''), where the potential profiles are typically smoother.

In this paper, we present an improved version of the model in Ref.~\citenum{Grassi13}, aiming at a better agreement with rigorous numerical quantum transport simulations. The main difference with respect to the previous work is the inclusion of a quantum rather than a semiclassical model to compute the transmission probability across potential steps. The new model is applied to both self-aligned GFETs and GFETs with gate underlaps. Its validity is assessed by comparing the terminal characteristics and internal quantities with those resulting from a self-consistent NEGF solver. In the case of GFETs with gate underlaps, the importance of a proper modeling of the electric field at the junctions will be highlighted.

We note that another semianalytical model for GFETs, accounting for the mode bottleneck effect and Klein tunneling, has been recently reported \cite{Alam13}. That model shares many similarities with ours. However, the expressions for charge and current included in our model have been rigorously derived in terms of the transmission probabilities at the junctions, rather than being empirically constructed. Moreover, in Ref.~\citenum{Alam13} the electrostatic problem is treated with a number of fitting parameters, whereas at most one fitting parameter is required in our model.

The paper is organized as follows. Sec.~\ref{sec_models} starts with an introduction to the two device structures under study, followed by a short description of the NEGF solver that is used for benchmarking the semianalytical model. The rest of that section is devoted to a detailed description of the equations that compose the semianalytical model. Simulation results are presented and discussed in Sec.~\ref{sec_results} and conclusions are finally drawn in Sec.~\ref{sec_conclusions}.

\section{\label{sec_models}Simulated devices and models}

\begin{figure}
\includegraphics[width=\columnwidth]{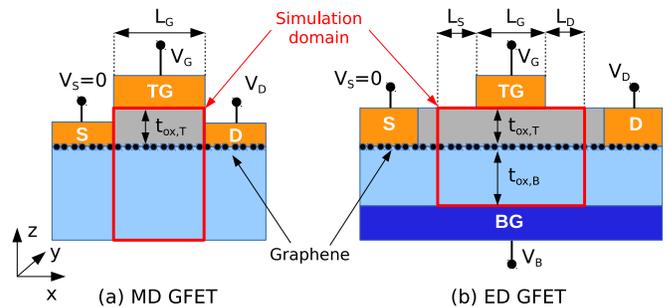}
\caption{\label{fig_devices}Longitudinal cross-sections of the two devices under study: (a) GFET with self-aligned contacts and thick back oxide; (b) GFET with top gate underlaps and back gate. S and D are the source and drain contacts, while TG and BG are the top and back gates, respectively. The source voltage is taken as the reference. The devices are considered to be infinite and homogeneous in the $y$-direction.}
\end{figure}

The schematics of the two device structures that we are going to model are shown in Fig.~\ref{fig_devices}. The first device has a thick back oxide and a gate self-aligned to the source and drain contacts (Fig.~\ref{fig_devices}a). The graphene under the source and drain contacts is doped by charge transfer, as a result of the difference of the metal and graphene workfunctions \cite{Giovannetti08}. The second device, instead, is a four-terminal structure: in addition to source, drain and gate, there is also a back gate terminal (Fig.~\ref{fig_devices}b). Moreover, gate underlaps are present at both sides of the channel. Such geometry allows the doping of the graphene in the underlap regions to be controlled electrostatically by the back gate voltage $V_B$. To stress the different doping mechanisms in the two devices, we name the first device ``metal-doped'' GFET (MD GFET) and the second one ``electrostatically doped'' GFET (ED GFET). 

The NEGF simulations are performed using an in-house developed code for GFETs, based on the self-consistent solution of the 2-D Poisson equation and the ballistic NEGF equations, with a $p_z$ tight-binding Hamiltonian. Taking advantage of the translational invariance in the $y$-direction, the 2-D transport problem is translated into a set of independent 1-D transport problems, one for each transverse wavevector $k_y$ \cite{Zhao11}. The rectangular simulation domains adopted for the two device structures are shown by red lines in Fig.~\ref{fig_devices}. For MD GFETs, the left/right edge of the simulation domain is placed at the interface between the top oxide and the source/drain contact. An ideal zero thickness of the source and drain contacts is assumed. Metal-induced doping is introduced by imposing Dirichlet boundary conditions in Poisson's equation at the vertical position corresponding to the graphene layer. In particular, the source/drain Dirac point energy $E_{d,S/D}$ is fixed at a distance $\Delta E_{\mathrm{con}}$ from the Fermi levels of the respective contacts. For ED GFETs, instead, the left and right edges of the simulation domain are placed inside the underlap regions. Here, Neumann rather than Dirichlet boundary conditions are imposed for the potential, implicitly assuming that the top gate and source and drain contacts are sufficiently separated so as to allow the electric potential to become almost $x$-independent inside the underlap regions ($x$ being the longitudinal or transport direction). For both devices, the source and drain self-energies are computed assuming semi-infinite leads, as in Ref.~\citenum{Zhao11} with the metal-graphene coupling strength $\Delta$ set to zero.

The semianalytical model is described in the following, starting from the electrostatics, then delineating the transport part, i.e., the equations for drain current and carrier concentrations as functions of the transmission probability across the potential steps, and at last specifying the model for the transmission probability itself.

\subsection{\label{sec_electrostatic}Electrostatic model}

\begin{figure}
\includegraphics[width=\columnwidth]{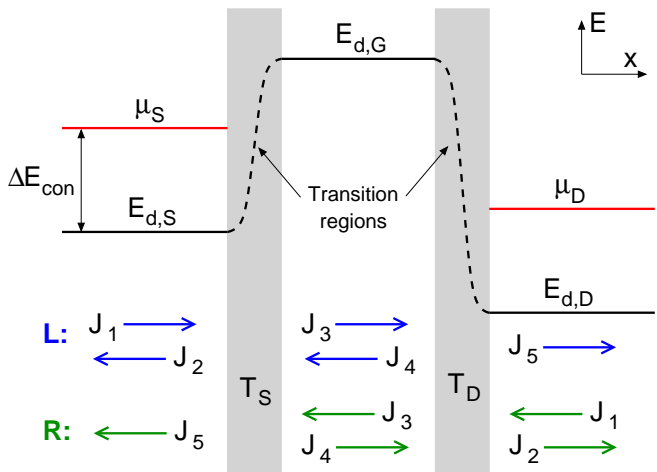}
\caption{\label{fig_diracpoint}Semianalytical model: schematic of the Dirac point profile $E_d(x)$ and pictorial representation of the current components $J_1 \ldots J_5(E,k_y)$ resulting from injection from source/drain (L/R) at fixed energy $E$ and $k_y$. To each transition region there corresponds a transmission probability, $T_S(E,k_y)$ or $T_D(E,k_y)$.}
\end{figure}

We consider a potential energy profile on the graphene layer as the one shown in Fig.~\ref{fig_diracpoint}. In particular, we assume that, in the source, drain, and channel regions sufficiently far from the source-channel and drain-channel interfaces, the potential can be approximated to be constant. The corresponding values of the Dirac point energy are denoted with $E_{d,S/D/G}$. The model applies to both devices. However, for the MD GFET, source and drain have to be identified with the graphene underneath the contacts, whereas, for the ED GFET, with the graphene in the underlap regions. The shape of the potential in the source/drain transition regions will be discussed in Sec.~\ref{sec_transmission}.

The computation of $E_{d,S/D/G}$ is different for the two devices. In the MD GFET, a fixed value independent of bias is imposed for the quantity $\Delta E_{\mathrm{con}}=\mu_{S/D}-E_{d,S/D}$ in order to emulate the metal-induced doping (see Fig.~\ref{fig_diracpoint}; $\mu_{S/D}$ is the source/drain Fermi level: hereafter, $\mu_S=0$). In the channel region, the Dirac point $E_{d,G}$ is self-consistently computed with the electron and hole densities (per unit area) $n_G$ and $p_G$, whose model will be described in Sec.~\ref{sec_transport}. More specifically, $E_{d,G}$ is computed from
\begin{equation}
q(n_G-p_G)=C_{\mathrm{ox},T}\left(V_G+E_{d,G}/q\right),
\label{eq_poisson_channel_MD}
\end{equation}
which corresponds to the solution of a 1-D Poisson equation in the vertical direction \cite{Grassi13}. Here, $V_G$ is the top gate voltage, $C_{\mathrm{ox},T}$ is the top oxide capacitance per unit area, and $q$ is the electron charge. A zero workfunction difference is assumed between top gate and graphene. In the ED GFET, the quantity $\mu_{S/D}-E_{d,S/D}$ is a function of bias. Therefore, an equation for $E_{d,S/D}$ analogue to (\ref{eq_poisson_channel_MD}) needs to be introduced:
\begin{equation}
q(n_{S/D}-p_{S/D})=C_{\mathrm{ox},B}\left(V_B+E_{d,S/D}/q\right),
\label{eq_poisson_source_drain_ED}
\end{equation}
where $n_{S/D}$ and $p_{S/D}$ are the electron and hole densities in source/drain (see Sec.~\ref{sec_transport}), $C_{\mathrm{ox},B}$ is the back oxide capacitance per unit area, and the workfunction difference between back gate and graphene is assumed zero. Moreover, the equation for $E_{d,G}$ has to be modified due to the presence of the back gate:
\begin{multline}
q(n_G-p_G)=C_{\mathrm{ox},T}\left(V_G+E_{d,G}/q\right)\\+C_{\mathrm{ox},B}\left(V_B+E_{d,G}/q\right).
\label{eq_poisson_channel_ED}
\end{multline}

\subsection{\label{sec_transport}Transport model}

We make the following assumptions about the propagation of electrons through the series of the two potential steps illustrated in Fig.~\ref{fig_diracpoint}: (\emph{i}) ballistic transport throughout the device, which implies conservation of total energy $E$ and transverse momentum (i.e., of $k_{y}$); (\emph{ii}) semiclassical transport in the flat potential regions of source, drain, and channel, which implies no interference between left-going and right-going particles.

The distribution of carriers inside the device can be obtained by superposition of the separate contributions due to injection from source and from drain. Let us consider for instance injection from source or left (L) at given $E$ and $k_y$, which gives rise to the current components $J_1 \ldots J_5$ represented with blue arrows in Fig.~\ref{fig_diracpoint}. Note that, from graphene dispersion relation $E(\vec{k})=E_d \pm \hbar v_f |\vec{k}|$ ($v_f$ is the graphene Fermi velocity, $\hbar=h/(2 \pi)$, and $h$ is Planck's constant), for a carrier to be propagating in the source (i.e., for $k_x$ to be real) it must be $|k_y|<|k_S|$ with
\begin{equation}
k_S=\frac{E-E_{d,S}}{\hbar v_f} .
\label{eq_kmax_source}
\end{equation}
Since transport is ballistic, carrier reflection occurs only in the regions where the potential varies (transition regions of Fig.~\ref{fig_diracpoint}). Let $T_S$ and $T_D$ be the transmission probabilities, dependent on $E$ and $k_y$, across the left and right transition regions, respectively. The model for them, based on a specific shape of the transition potential, will be presented in Sec.~\ref{sec_transmission}. Once $T_{S}$ and $T_{D}$ are known, it is possible to calculate $J_2 \ldots J_5$ as functions of $J_1$. Under the assumption of semiclassical transport inside the channel, the following expressions can be derived:
\begin{align}
\left(\frac{J_2}{J_1}\right)_{\!\!L}&=\frac{T_S(1-T_D)+T_D(1-T_S)}{1-(1-T_S)(1-T_D)}, \label{eq_J2} \\
\left(\frac{J_3}{J_1}\right)_{\!\!L}&=\frac{T_S}{1-(1-T_S)(1-T_D)}, \label{eq_J3} \\
\left(\frac{J_4}{J_1}\right)_{\!\!L}&=\frac{T_S(1-T_D)}{1-(1-T_S)(1-T_D)}, \label{eq_J4} \\
\left(\frac{J_5}{J_1}\right)_{\!\!L}&\equiv T=\frac{T_S T_D}{1-(1-T_S)(1-T_D)} \label{eq_J5},
\end{align}
where the subscript L indicates injection from left and $T$ is the total transmission probability from source to drain. Injection from drain or right (R, see Fig.~\ref{fig_diracpoint}) results in identical expressions, except for the interchange of $T_S$ and $T_D$.

The drain current and carrier concentrations can be expressed in terms of the coefficients in (\ref{eq_J2})--(\ref{eq_J5}). In order to do so, the following considerations must be made regarding the contributions at given $E$ and $k_y$ (we refer for instance to the case of injection from source): (\emph{i}) $J_1 \propto f_S(E)$ for electrons, with $f_S(E)$ the contact Fermi distribution with Fermi level $\mu_S$; (\emph{ii}) $J_1 \propto 1-f_S(E)$ for holes; (\emph{iii}) the drain current can be evaluated in any of the three flat potential regions of Fig.~\ref{fig_diracpoint} as the difference of the currents of left-going and right-going particles; (\emph{iv}) the carrier concentrations in each of three flat potential regions can be computed as the sum of the currents of left-going and right-going particles divided by $q |v_x|$, where $v_x = (1/\hbar)\partial E(\vec{k})/\partial k_x$ is the longitudinal carrier velocity. For the drain current (per unit width) $I$, we then have
\begin{align}
I&=I_L-I_R , \label{eq_current} \\
I_{L/R}&=\frac{4q}{\pi h}\int_{-\infty}^{\infty}dE\int_{0}^{|k_{S/D}|}dk_{y}T(E,k_{y})\,f_{S/D}(E) ,
\label{eq_current_left}
\end{align}
with obvious definitions of $k_D$ and $f_D(E)$. In (\ref{eq_current_left}) the factor of 4 is due to spin and valley degeneracies and the current is computed from the filled states (electrons) of both the conduction and valence bands. Noting that it is necessarily $T=0$ for $k_{y}>\min\{|k_S|,|k_D|\}$ and defining the transmission function (per unit width) $\overline{T}(E)$ as \cite{DattaQT2005}
\begin{equation}
\overline{T}(E)=\frac{2}{\pi}\int_{0}^{\infty}dk_{y}\,T(E,k_{y}) ,
\label{eq_transmissionfunction}
\end{equation}
(\ref{eq_current}) can be recast in the well-known Landauer formula
\begin{equation}
I=\frac{2q}{h}\int_{-\infty}^{\infty}dE\, \overline{T}(E) [f_{S}(E)-f_{D}(E)] .
\label{eq_current_landauer}
\end{equation}
As far as the electron density in the channel is concerned, we have
\begin{align}
n_G&= n_G^L+n_G^R , \label{eq_electron_channel} \\
n_{G}^{L/R}&=\int_{E_{d,G}}^{\infty}dE\left[\frac{4}{\pi h}\int_{0}^{|k_{S/D}|}dk_{y} \left(\frac{J_{3}+J_{4}}{J_{1}}\right)_{\!\!L/R}\frac{1}{|v_{x}^{G}|}\right] \nonumber \\
&\hphantom{=} \times f_{S/D}(E) . \label{eq_electron_channel_left}
\end{align}
Here, $v_{x}^{G}$ is the longitudinal carrier velocity evaluated in the channel region. The quantity in square brackets can be understood as the DOS in the channel related to injection from source/drain. The expression of the hole density $p_G$ can be obtained from (\ref{eq_electron_channel}) and (\ref{eq_electron_channel_left}) replacing $f_{S/D}(E)$ with $1-f_{S/D}(E)$ and letting the energy integrals go from $-\infty$ to $E_{d,G}$. For the MD GFET, (\ref{eq_electron_channel}) and (\ref{eq_electron_channel_left}) and the corresponding equations for $p_G$ are solved self-consistently with (\ref{eq_poisson_channel_MD}) for $E_{d,G}$. For the ED GFET, one also needs the expressions of $n_{S/D}$ and $p_{S/D}$. The set of equations for $n_{S/D/G}$ and $p_{S/D/G}$ are solved self-consistently with (\ref{eq_poisson_source_drain_ED}) and (\ref{eq_poisson_channel_ED}) for $E_{d,S/D/G}$. We give for example the expression of $n_S$, the other ones being straightforward to derive by analogy:
\begin{align}
n_S&= n_S^L+n_S^R , \label{eq_electron_source} \\
n_{S}^{L}&=\int_{E_{d,S}}^{\infty}dE\left[\frac{4}{\pi h}\int_{0}^{|k_S|}dk_{y}\left(1+\frac{J_{2}}{J_{1}}\right)_{\!\!L}\frac{1}{|v_{x}^{S}|}\right] \nonumber\\
&\hphantom{=}\times f_{S}(E) , \label{eq_electron_source_left} \\
n_{S}^{R}&=\int_{E_{d,S}}^{\infty}dE\left[\frac{4}{\pi h}\int_{0}^{|k_D|}dk_{y}\left(\frac{J_{5}}{J_{1}}\right)_{\!\!R}\frac{1}{|v_{x}^{S}|}\right] \nonumber\\
&\hphantom{=}\times f_{D}(E) . \label{eq_electron_source_right}
\end{align}
It should be noted that all the integrals appearing in (\ref{eq_current_left}), (\ref{eq_electron_channel_left}), (\ref{eq_electron_source_left}), and (\ref{eq_electron_source_right}), not only the ones over $E$ but also the ones over $k_y$, have to be performed numerically since the expressions of the transmission probabilities (see Sec.~\ref{sec_transmission}) do not allow in general analytical solutions. The singularity at $k_y=0$ of the $k_y$ integrands in (\ref{eq_electron_channel_left}), (\ref{eq_electron_source_left}), and (\ref{eq_electron_source_right}) can be eliminated with the change of variables $k_y \rightarrow \theta = \arctan(k_y/k_x)$.

\subsection{\label{sec_transmission}Transmission model}

\begin{figure}
\includegraphics[scale=0.32]{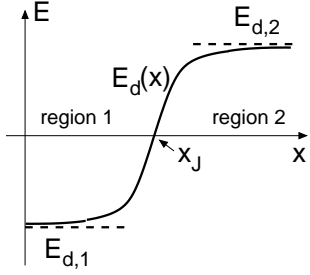}
\caption{\label{fig_step}Schematic representation of the potential energy step.}
\end{figure}

In order to complete the model we have to introduce the equations for $T_{S}$ and $T_{D}$. Let ``region 1'' and ``region 2'' be the two regions adjacent to the generic transition region and $E_{d,1}$ and $E_{d,2}$ the respective Dirac point energies as shown in Fig.~\ref{fig_step}. The transition probability across the transition region will be denoted as $T_{\mathrm{step}}$.

Following Ref.~\citenum{Cayssol09}, we model the profile of the potential energy step as
\begin{equation}
E_{d}(x)=E_{d,1}+\frac{E_{d,2}-E_{d,1}}{e^{-(x-x_J)/d}+1} ,
\label{eq_smoot_potential}
\end{equation}
where $x_J$ is the midpoint of the transition region. According to this formula, the values $E_{d,1}$ and $E_{d,2}$ are reached only asymptotically as illustrated in Fig.~\ref{fig_step}. The transition length $d$ is a parameter which can be related to the maximum electric field value $F=(1/q)\left|dE_{d}/dx\right|_{x=x_J}$ through
\begin{equation}
d=\frac{|E_{d,2}-E_{d,1}|}{4qF} .
\label{eq_field_distance}
\end{equation}
The limit $d \rightarrow 0$ ($F \rightarrow \infty$) corresponds to an abrupt junction. 

Using the single-valley Dirac Hamiltonian, the quantum mechanical problem of electron scattering through the potential profile in (\ref{eq_smoot_potential}) can be solved analytically \cite{Cayssol09}. The expression of the transmission probability evaluated asymptotically far from the junction is
\begin{equation}
T_{\mathrm{step}}=\vartheta(\min\{|k_1|,|k_2|\}-|k_y|) T_{\mathrm{qu}}
\label{eq_smooth_transmission}
\end{equation}
with
\begin{equation}
T_{\mathrm{qu}} = 1-\frac{\sinh(\pi d \kappa^{+-})\sinh(\pi d \kappa^{-+})}
{\sinh(\pi d \kappa^{++})\sinh(\pi d \kappa^{--})} .
\label{eq_quantum_transmission}
\end{equation}
The symbols are defined as follows: $\vartheta$ is the Heaviside step function, $\kappa^{\rho \sigma}=k_1-k_2+\rho k_{x,1}+\sigma k_{x,2}$ ($\rho,\sigma=\pm 1$), and
\begin{align}
k_{m}&=\frac{E-E_{d,m}}{\hbar v_{f}} , \label{eq_k}\\
k_{x,m}&=\mathrm{sgn}(k_{m})\sqrt{k_{m}^{2}-k_{y}^{2}} , \label{eq_kx}
\end{align}
with $m=1,2$. Definition (\ref{eq_k}) is the same as (\ref{eq_kmax_source}). The $\vartheta$ function in (\ref{eq_smooth_transmission}) accounts for the phenomenon of total reflection, responsible for the mode bottleneck effect mentioned before: for an incident electron with energy $E$ and transverse momentum $k_y$, transmission is forbidden if propagating states with the same $E$ and $k_y$ are not available on the other side of the junction. The factor $T_{\mathrm{qu}}$, which is of quantum mechanical nature, gives partial reflection even if those states are available and is dependent on the parameter $d$. For an abrupt junction ($d\rightarrow 0$), we have
\begin{equation}
T_{\mathrm{qu}} \rightarrow \frac{2k_{x,1}k_{x,2}}{k_{1}k_{2}+k_{x,1}k_{x,2}-k_{y}^{2}} ,
\label{eq_quantum_abrupt_transmission}
\end{equation}
which is the same result one can obtain by taking the wavefunctions of each region equal to the eigenfunctions of the free electron Dirac Hamiltonian and requiring them to be continuous at the interface \cite{Low12}. The fact that, even for an abrupt junction, transmission can be lower than one represents the effect of wavefunction mismatch that was mentioned previously. In Sec.~\ref{sec_results} we will show that the limit of abrupt junction works remarkably well for MD GFETs, whereas for ED GETs one needs to complete the model with an equation for $d$ (or $F$) as a function of the device parameters and of bias. Finally, it is worth noticing that, if one sets $T_{\mathrm{qu}}=1$ in (\ref{eq_smooth_transmission}) and substitutes the resulting expressions of $T_S$ and $T_D$ in (\ref{eq_J2})--(\ref{eq_J5}), the $k_y$ integrals in (\ref{eq_current_left}), (\ref{eq_electron_channel_left}), (\ref{eq_electron_source_left}), and (\ref{eq_electron_source_right}) can be calculated analytically. Doing this for the MD GFET, one arrives at the semianalytical semiclassical model in Ref.~\citenum{Grassi13}.

\section{\label{sec_results}Results}

We start considering the MD GFET (Fig.~\ref{fig_devices}a). The simulated device has $\Delta E_{\mathrm{con}} = 0.4$~eV, top oxide thickness $t_{\mathrm{ox},T} = 0.5$~nm, and a gate length $L_G=50$~nm. Top and back dielectrics are both made of silicon oxide. Note that the chosen value of $\Delta E_{\mathrm{con}}$ corresponds to n-type doped source and drain regions. Depending on the sign of $V_G$, an n-p-n or n-n-n double junction is created inside the device resulting respectively in NDR or quasi-saturation regime \cite{Grassi13}.

\begin{figure}
\includegraphics[width=\columnwidth]{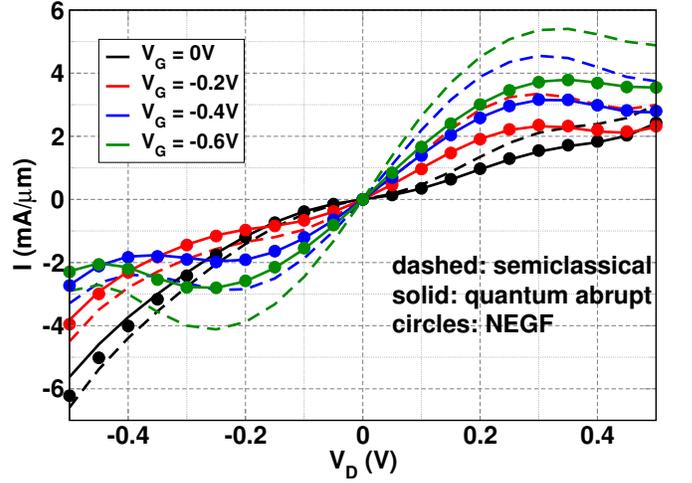}
\caption{\label{fig_I_Vds_ndr_MDGFET}Output characteristics of the MD GFET in the NDR regime. Comparison between semianalytical semiclassical model ($T_{\mathrm{qu}}=1$), semianalytical quantum model with abrupt junctions ($T_{\mathrm{qu}}$ from Eq.~\ref{eq_quantum_abrupt_transmission}), and NEGF.}
\end{figure}

In Fig.~\ref{fig_I_Vds_ndr_MDGFET}, we plot the output characteristics for $V_G\le 0$ (p-type channel) obtained with the following models: (\emph{i}) semianalytical model with $T_{\mathrm{qu}}=1$ (i.e., semiclassical model of Ref.~\citenum{Grassi13}), (\emph{ii}) semianalytical model with $T_{\mathrm{qu}}$ computed according to the limit of abrupt junction in (\ref{eq_quantum_abrupt_transmission}), and (\emph{iii}) NEGF $+$ 2-D electrostatics. It can be seen that the use of a quantum rather than a semiclassical model of the transmission probability greatly improves the accuracy of the semianalytical model resulting in $I$--$V$ curves almost perfectly overlapping with the NEGF ones (error $<10 \%$).
\begin{figure}
\includegraphics[width=\columnwidth]{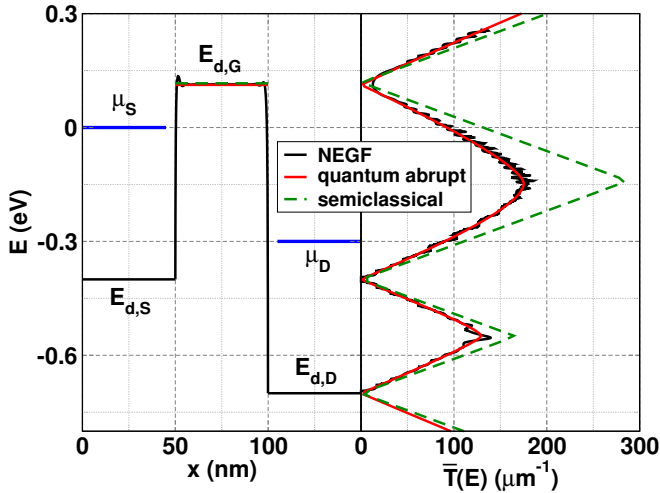}
\caption{\label{fig_M_Vg-0.4_Vd0.3_MDGFET}Dirac point energy profile (left) and transmission function vs.~energy (right) of the MD GFET at $V_G=-0.4$~V and $V_D=0.3$~V. Same models as in Fig.~\ref{fig_I_Vds_ndr_MDGFET}.}
\end{figure}
In Fig.~\ref{fig_M_Vg-0.4_Vd0.3_MDGFET}, we show the band profile $E_d(x)$ and the transmission function obtained with the three models at a bias at the edge of the NDR region at positive $V_D$. The shape of the potential profile resulting from the NEGF simulation demonstrates that the approximation of abrupt junctions is very well verified. This is due to the ideal geometry that we have considered, with self-aligned contacts and zero thickness source and drain contacts. All three models give essentially the same value of the mid-channel Dirac point $E_{d,G}$. However, the semianalytical semiclassical model largely overestimates $\overline{T}(E)$, especially in the energy window where double Klein tunneling occurs (that is the one between $E_{d,S}$ and $E_{d,G}$, which, in this case, contains the energy range between $\mu_S$ and $\mu_D$ that contributes most to current), due to the aforementioned neglect of wavefunction mismatch at the junctions. On the other hand, the semianalytical quantum model reproduces closely the transmission function from NEGF, except for minimal differences (absence of resonance peaks and absence of the direct source-to-drain tunneling contribution for energies around $E_{d,G}$) related to the assumption of semiclassical transport inside the channel.
\begin{figure}
\includegraphics[width=\columnwidth]{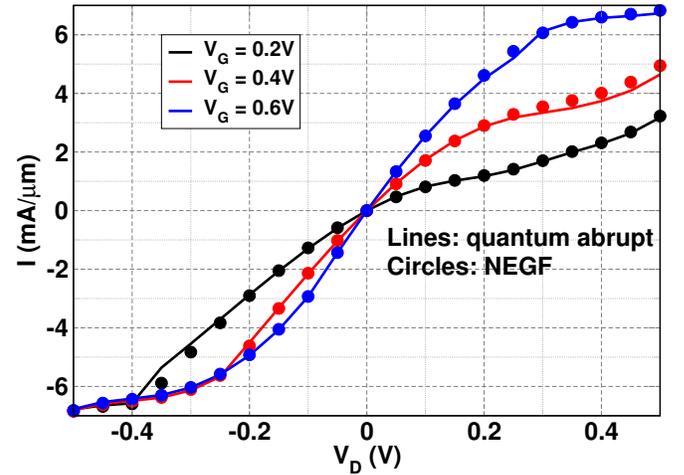}
\caption{\label{fig_I_Vds_sat_MDGFET}Output characteristics of the MD GFET in the quasi-saturation regime. Comparison between semianalytical quantum model with abrupt junctions and NEGF.}
\end{figure}
The agreement between these two models is remarkably good also in the quasi-saturation regime, as demonstrated by Fig.~\ref{fig_I_Vds_sat_MDGFET}, where the output characteristics for $V_G>0$ (n-type channel) are shown. 

Let us then move to the ED GFET (Fig.~\ref{fig_devices}b). The gate length is $L_G=50$~nm and the lengths of the source and drain extensions are $L_S=L_D=15$~nm. The top dielectric is Al$_{2}$O$_{3}$, while the back dielectric is silicon oxide with thickness $t_{\mathrm{ox},B}=10$~nm. The top oxide thickness will be treated as a parameter: the reference device has $t_{\mathrm{ox},T}=$1.2~nm (effective oxide thickness $\mathrm{EOT} = 0.5$~nm). The back gate voltage is held fixed at $V_B=9$~V, yielding a heavy n-type doping of the source and drain regions.

\begin{figure}
\includegraphics[width=\columnwidth]{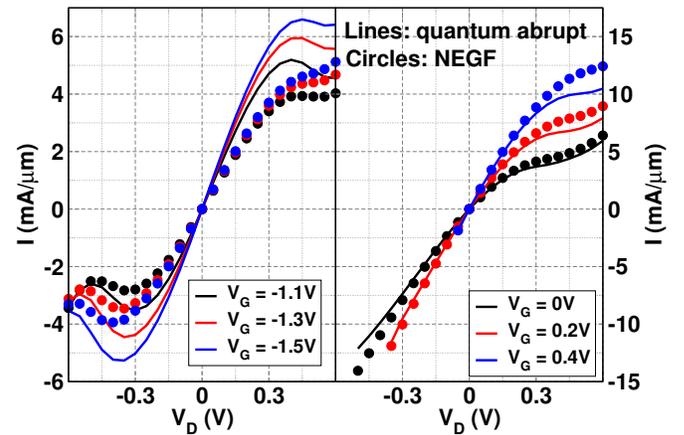}
\caption{\label{fig_I_Vds_EDGFET_abrupt}Output characteristics of the reference ED GFET in the NDR (left) and quasi-saturation (right) regimes. Comparison between semianalytical quantum model with abrupt junctions and NEGF. The curves for $V_G=0.4$~V and negative $V_D$ are not drawn due to convergence problems of the NEGF code.}
\end{figure}

\begin{figure}
\includegraphics[width=\columnwidth]{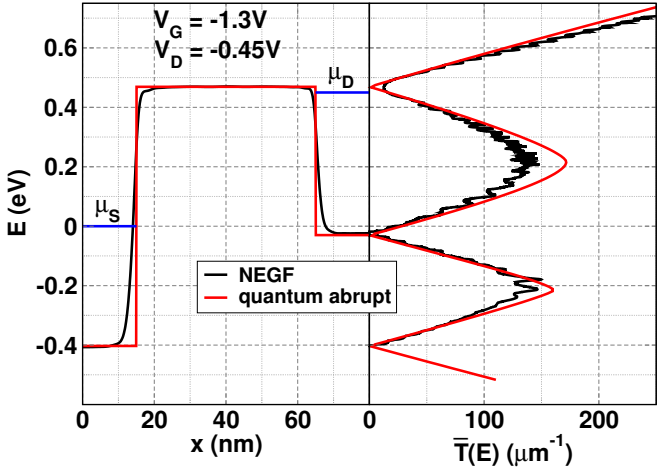}
\caption{\label{fig_M_Vg-1.3_Vd-0.45_EDGFET_abrupt}Dirac point energy profile (left) and transmission function vs.~energy (right) of the reference ED GFET at $V_G=-1.3$~V and $V_D=-0.45$~V. Same models as in Fig.~\ref{fig_I_Vds_EDGFET_abrupt}.}
\end{figure}

\begin{figure}
\includegraphics[width=\columnwidth]{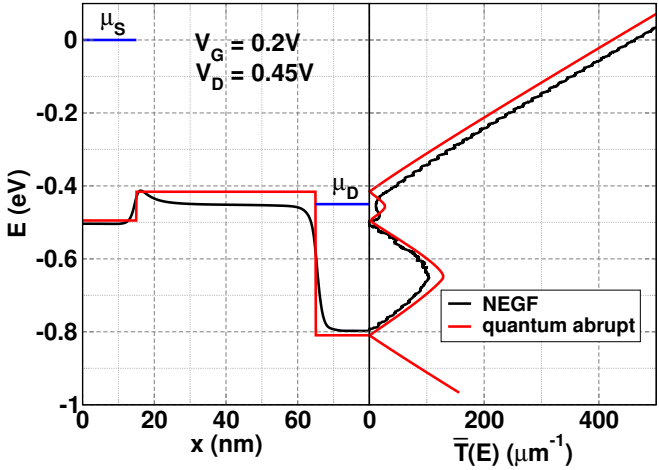}
\caption{\label{fig_M_Vg0.2_Vd0.45_EDGFET_abrupt}Same as in Fig.~\ref{fig_M_Vg-1.3_Vd-0.45_EDGFET_abrupt} but at $V_G=0.2$~V and $V_D=0.45$~V.}
\end{figure}

In Fig.~\ref{fig_I_Vds_EDGFET_abrupt}, the current vs.~drain voltage characteristics in the NDR and quasi-saturation regimes of the reference ED GFET are shown. In this case, the agreement between the semianalytical quantum model assuming abrupt junctions and NEGF is only qualitative. In order to understand the reason, we look again at the potential energy profiles and transmission functions at some selected biases (Figs.~\ref{fig_M_Vg-1.3_Vd-0.45_EDGFET_abrupt} and \ref{fig_M_Vg0.2_Vd0.45_EDGFET_abrupt}). 
At the bias of Fig.~\ref{fig_M_Vg-1.3_Vd-0.45_EDGFET_abrupt}, which is inside the NDR region at negative $V_D$, the potential computed with NEGF reaches, in the source, drain, and gate regions far from the junctions, values very similar to the ones computed with the semianalytical model. However, at energies corresponding to double or single Klein tunneling, the NEGF simulation predicts values of $\overline{T}(E)$ somewhat lower than the semianalytical model. This has to do with the shape of the potential transitions, which are smoother than in MD GFETs and force electrons incident at non-normal angles ($k_y \ne 0$) to tunnel through an apparent band gap, as mentioned in Sec.~\ref{sec_intro}. 
The situation is different in the quasi-saturation region (Fig.~\ref{fig_M_Vg0.2_Vd0.45_EDGFET_abrupt}). In the energy window between $\mu_S$ and $\mu_D$, transport is over-the-barrier and thus the shape of the potential at the junctions is expected to have a lesser impact than in the NDR case. Nevertheless, the semianalytical model seems to wrongly estimate the level of the channel potential, causing the transmission function to shift up in energy compared to the NEGF result. This might be related to the ``bump'' seen in the NEGF potential, which causes double band-to-band tunneling at the source-channel interface and cannot be captured by the semianalytical model. 

The results shown above demonstrate the need to properly model the electric field in ED GFETs, going beyond the limit of abrupt junction. In order to do that, we follow the screening theory in Ref.~\citenum{Zhang08}. Referring again to the notation in Fig.~\ref{fig_step}, let ``region 1'' be the top-gated region (i.e., the channel region) of each junction. We denote the net electron density by $\rho(x)=n(x)-p(x)$ and its asymptotic values far from the junction by $\rho_1$ and $\rho_2$. We distinguish two cases depending on the relative sign of $\rho_1$ and $\rho_2$. If $\rho_1 \rho_2 < 0$ (p-n junction), we use the model in Ref.~\citenum{Zhang08}, which takes into account the weak screening effect due to $\rho(x)$ going to zero at the junction and expresses the maximum electric field at the junction as
\begin{equation}
q F=\frac{1}{0.186} \times \hbar v_{f}\alpha^{1/3}\left|\frac{\rho_{1}}{t_{\mathrm{ox},T}}\right|^{2/3}\left(1-\frac{\rho_{1}}{\rho_{2}}\right)^{-4/3} ,
\label{eq_field_pn}
\end{equation}
where $\alpha=q^2/(4 \pi \epsilon_{\mathrm{ox},T} \hbar v_f)$ and $\epsilon_{\mathrm{ox},T}$ is the top oxide dielectric constant ($\epsilon_{\mathrm{ox},T}/t_{\mathrm{ox},T}=C_{\mathrm{ox},T}$). If $\rho_1 \rho_2 > 0$ (n-n or p-p junction), we use the expression
\begin{align}
q F&=\frac{\pi^{3/2} \hbar v_f}{2 t_{\mathrm{ox},T}} \frac{r}{(1+r)^{5/2}} \frac{|\rho_1-\rho_2|}{|\rho_1+r\rho_2|^{1/2}} ,
\label{eq_field_nn} \\
r&=\frac{1-3\rho_1/\rho_2+\sqrt{\left(1-3\rho_1/\rho_2\right)^2+32\rho_1/\rho_2}}{8} \label{eq_r} ,
\end{align}
which we have derived according to Ref.~\citenum{Zhang08} but assuming nearly perfect screening. See Appendix~\ref{sec_field_nn} for details. Eqs.~\ref{eq_field_pn}--\ref{eq_r} provide, for each junction, the value of $F$ to insert into (\ref{eq_field_distance}) to compute $d$ and thus $T_S$ or $T_D$. Due to the dependence on the net electron density, (\ref{eq_field_pn})--(\ref{eq_r}) need to be solved self-consistently with the transport equations (\ref{eq_electron_channel})--(\ref{eq_electron_source_right}).

\begin{figure}
\includegraphics[width=\columnwidth]{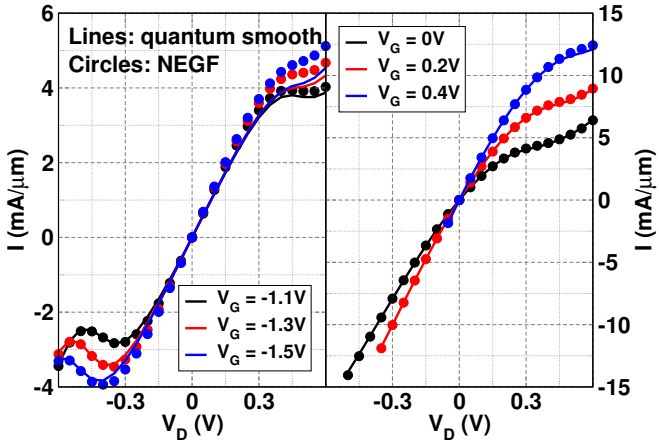}
\caption{\label{fig_I_Vds_EDGFET_smooth}Output characteristics of the reference ED GFET in the NDR (left) and quasi-saturation (right) regimes. Comparison between semianalytical quantum model with smooth junctions ($T_{\mathrm{qu}}$ from Eq.~\ref{eq_quantum_transmission}, with $d$ from Eq.~\ref{eq_field_distance} and $F$ from Eqs.~\ref{eq_field_pn}--\ref{eq_r}) and NEGF.}
\end{figure}

\begin{figure}
\includegraphics[width=\columnwidth]{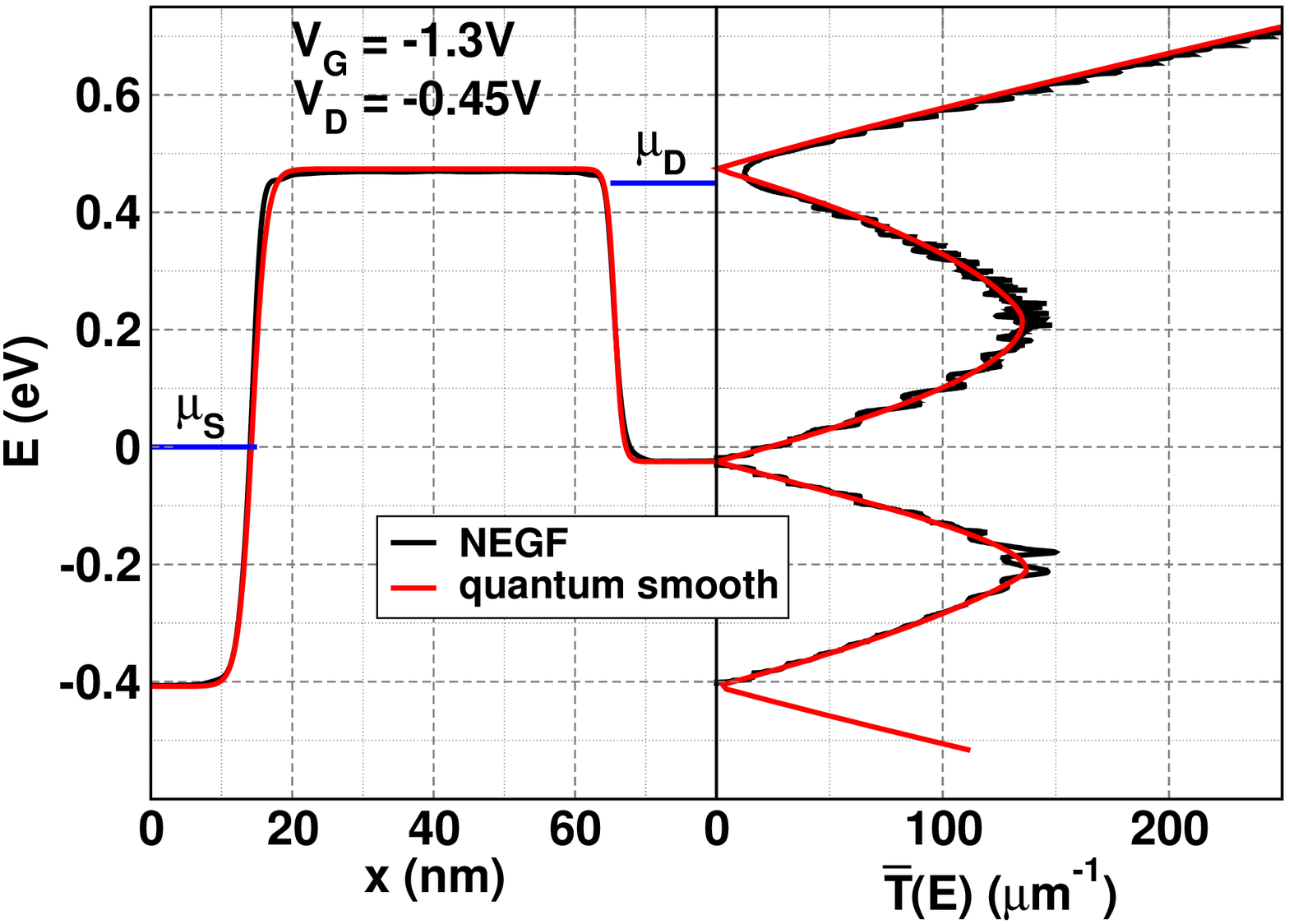}
\caption{\label{fig_M_Vg-1.3_Vd-0.45_EDGFET_smooth}Same as in Fig.~\ref{fig_M_Vg-1.3_Vd-0.45_EDGFET_abrupt} but using the semianalytical quantum model with smooth junctions.}
\end{figure}

\begin{figure}
\includegraphics[width=\columnwidth]{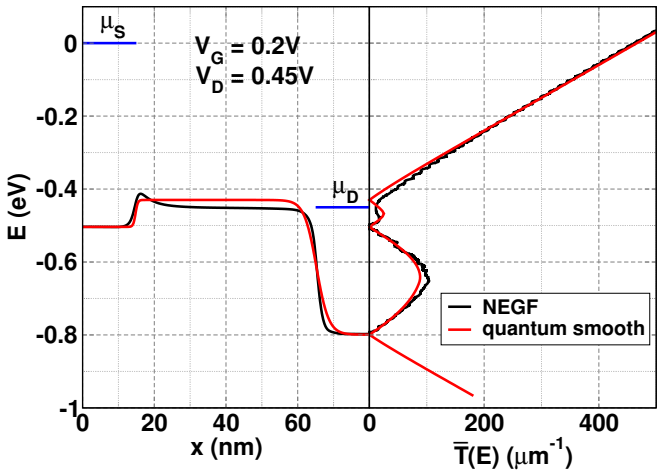}
\caption{\label{fig_M_Vg0.2_Vd0.45_EDGFET_smooth}Same as in Fig.~\ref{fig_M_Vg0.2_Vd0.45_EDGFET_abrupt} but using the semianalytical quantum model with smooth junctions.}
\end{figure}

The current vs.~drain voltage characteristics obtained with the improved semianalytical model accounting for smooth junctions are plotted in Fig.~\ref{fig_I_Vds_EDGFET_smooth}, superimposed to the NEGF curves, which are repeated from Fig.~\ref{fig_I_Vds_EDGFET_abrupt}. The accuracy of the semianalytical model is greatly improved as compared to Fig.~\ref{fig_I_Vds_EDGFET_abrupt}. The agreement with NEGF is now made quantitative both in the NDR regime at negative $V_D$, with only some difference in the predicted output conductance at low drain bias (error $< 15 \%$), and, somewhat surprisingly, in the quasi-saturation regime, where minimal deviations are only seen at high drain and gate biases (error $< 3 \%$). A larger error ($\approx 13 \%$) is observed in the NDR regime at positive $V_D$. The internal quantities (band profile and transmission function) are compared in Fig.~\ref{fig_M_Vg-1.3_Vd-0.45_EDGFET_smooth} and Fig.~\ref{fig_M_Vg0.2_Vd0.45_EDGFET_smooth}, which are obtained at the same bias points of Fig.~\ref{fig_M_Vg-1.3_Vd-0.45_EDGFET_abrupt} and Fig.~\ref{fig_M_Vg0.2_Vd0.45_EDGFET_abrupt}, respectively. The Dirac point profile of the semianalytical model is constructed as follows: (\emph{i}) for $x<40$~nm, we plot (\ref{eq_smoot_potential}) with parameters taken from the source and channel regions; (\emph{ii}) for $x>40$~nm, we plot (\ref{eq_smoot_potential}) with parameters taken from the drain and channel regions; (\emph{iii}) we compute $x_J$ as $x_J=x_G \mp (t_{\mathrm{ox},T}/\pi)(\ln r + 1 + r)$, where $x_G$ is the position of the left/right edge of the top gate and the parameter $r$ is computed as $r=-\rho_1/\rho_2$ if $\rho_1 \rho_2 < 0$ (see Ref.~\citenum{Zhang08}) or according to (\ref{eq_r}) if $\rho_1 \rho_2 > 0$ (see Appendix~\ref{sec_field_nn}). Looking at Fig.~\ref{fig_M_Vg-1.3_Vd-0.45_EDGFET_smooth}, it is seen that, in the NDR region at negative $V_D$, the potential shape given by (\ref{eq_smoot_potential}) with the electric field model in (\ref{eq_field_pn})--(\ref{eq_r}) reproduces with very good accuracy the potential profile from the NEGF simulation, so that the transmission functions from the semianalytical model and NEGF are almost perfectly overlapping. The electric field model is less accurate in the NDR region at positive $V_D$ (not shown). In the quasi-saturation region (Fig.~\ref{fig_M_Vg0.2_Vd0.45_EDGFET_smooth}), the mid-channel Dirac point level of the semianalytical model seems still to be far from the NEGF result. However, such value of $E_{d,G}$ allows the transmission functions from the semianalytical model and NEGF to match closely for $E>E_{d,G}$, compensating the effect of the potential bump.

\begin{figure}
\includegraphics[width=\columnwidth]{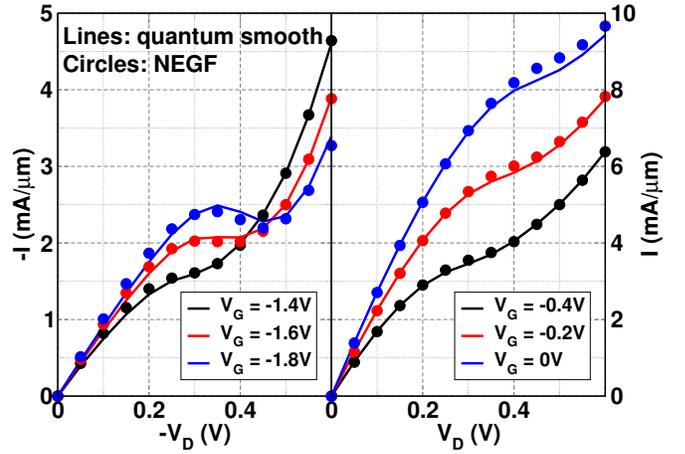}
\caption{\label{fig_I_Vds_EDGFET_smooth_tox2.4}Output characteristics of the ED GFET with thicker top oxide ($t_{\mathrm{ox},T}=2.4$~nm) in the NDR (left) and quasi-saturation (right) regimes. Comparison between semianalytical quantum model with smooth junctions and NEGF.}
\end{figure}

We conclude by showing in Fig.~\ref{fig_I_Vds_EDGFET_smooth_tox2.4} the output characteristics obtained by doubling the top oxide thickness ($t_{\mathrm{ox},T}=2.4$~nm or $\mathrm{EOT}=1$~nm). The accuracy of the semianalytical model is still very good despite the increased fringing effect induced by the larger $t_{\mathrm{ox},T}$, further confirming the validity of the electric field model in (\ref{eq_field_pn})--(\ref{eq_r}).

\section{\label{sec_conclusions}Conclusions}

In this work, a semianalytical model for short-channel GFETs has been presented, which improves the one in Ref.~\citenum{Grassi13}. The model applies to two types of devices, MD GFETs and ED GFETs, which differ in the doping mechanism (metal-induced vs.~electrostatic) of the source and drain regions. Charge and current are expressed in terms of the transmission probabilities across the source-channel and drain-channel junctions. Such probabilities are computed through a quantum model based on a specific shape of the potential transition, which can be tuned to describe both abrupt and smooth junctions. The semianalytical model has been benchmarked against self-consistent NEGF simulations. For MD GFETs, the semianalytical model using the approximation of abrupt junctions reproduces almost perfectly the $I$--$V$ characteristics computed with NEGF, both in the bias region of quasi-saturation and in the one of NDR, demonstrating, unlike the previous model in Ref.~\citenum{Grassi13}, an accurate description of Klein tunneling. For ED GFETs, the limit of abrupt junction does not work as well. However, if completed with a model of the electric field at the junctions, the semianalytical model can restore a good quantitative agreement with NEGF, even in devices with thicker top oxide layers.

Due to the different implementations (MATLAB vs. Fortran languages), it has not been possible to fairly measure the speed-up of the semianalytical model compared to the NEGF solver. Although the complexity of the semianalytical model has risen compared to the previous work in Ref.~\citenum{Grassi13} due to the additional presence of numerical integrals over $k_y$, we believe the semianalytical model to still maintain a sizable computational advantage over NEGF.

\appendix

\section{Derivation of the electric field model in Eqs.~\ref{eq_field_nn} and \ref{eq_r}} \label{sec_field_nn}

\begin{figure}
\includegraphics[scale=0.32]{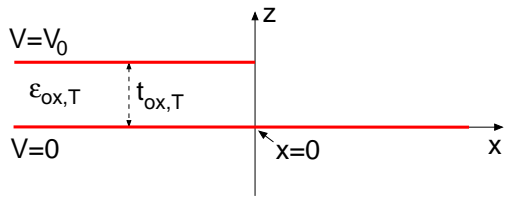}
\caption{\label{fig_screening_capacitor}Semi-infinite top gate over infinite graphene sheet, treated as an ideal metallic layer.}
\end{figure}

In order to find an approximate expression for the maximum electric field of a unipolar (i.e., n-n or p-p) junction, we rely on the Thomas-Fermi approximation described in Ref.~\citenum{Zhang08}. We refer to the geometry illustrated in Fig.~\ref{fig_screening_capacitor}, where the top gate and the graphene sheet are treated as semi-infinite and infinite layers, respectively. The origin of the $x$-axis is placed at the edge of the top gate. The device is at equilibrium. Underneath the graphene layer there is the infinite back gate, not shown in the figure. Let $V_0$ be the voltage difference between top gate and graphene for $x \rightarrow -\infty$. At $x\rightarrow \infty$ the graphene carrier density $\rho$ tends to the value $\rho_2$, which is fixed by the back gate voltage. At $x \rightarrow -\infty$, $\rho \rightarrow \rho_1$, with $\rho_1-\rho_2=C_{\mathrm{ox},T}V_0/q$.

If graphene were an ideal metal, the potential on the graphene layer $V(x)=-E_d(x)/q$ would be constant and the corresponding carrier density $\rho(x)$ could be calculated by solving the parallel plate capacitor electrostatic problem in Fig.~\ref{fig_screening_capacitor}. This can be done analytically and yields in parametric form \cite{Morse53}:
\begin{equation}
\begin{cases}
\rho=\frac{\rho_1-\rho_2}{1+r}+\rho_2=\rho(r), \quad 0<r<\infty , \\
x=\frac{t_{\mathrm{ox},T}}{\pi}\left[\ln r + 1 + r\right]=x(r) .
\end{cases}
\label{eq_charge_capacitor}
\end{equation}
In reality, graphene has a finite DOS, so that the actual carrier density differs from (\ref{eq_charge_capacitor}) and the potential is not identically zero. However, if screening is nearly perfect, (\ref{eq_charge_capacitor}) is a good approximation of the carrier density and a first-order approximation of the potential can be computed from the relation between $E_d$ and $\rho$ in graphene at equilibrium. The latter, in the zero temperature limit, takes the form
\begin{equation}
E_d(x)=-\mathrm{sgn}(\rho(x))\sqrt{\pi}\hbar v_{f} |\rho(x)|^{1/2},
\label{eq_chemical_potential}
\end{equation}
where energies are measured with respect to the Fermi level. Combining (\ref{eq_charge_capacitor}) with (\ref{eq_chemical_potential}), we get
\begin{align}
\frac{d E_d}{dx}&=-\frac{\sqrt{\pi}\hbar v_{f}}{2} \left|\frac{\rho_1-\rho_2}{1+r}+\rho_2\right|^{-1/2} \frac{d\rho}{dr} \frac{dr}{dx} \nonumber \\
&=\frac{\pi^{3/2} \hbar v_f}{2 t_{\mathrm{ox},T}} \frac{r}{(1+r)^{5/2}} \frac{\rho_1-\rho_2}{\left|\rho_1+r\rho_2\right|^{1/2}}.
\label{eq_derivative_potential}
\end{align}
Taking the absolute value of (\ref{eq_derivative_potential}) gives (\ref{eq_field_nn}), where the coordinate $r$ still needs to be fixed. This can be done by requiring the electric field to be maximum, i.e., by taking the derivative of (\ref{eq_derivative_potential}) with respect to $r$ and equating it to zero. This gives a quadratic equation in $r$ whose only physically acceptable solution ($r>0$) is the one in (\ref{eq_r}).



%
%

%

\begin{acknowledgments}
The authors would like to thank Dr.~T.~Low of IBM T. J. Watson Research Center for useful discussions. This work was supported by the EU Project GRADE 317839.
\end{acknowledgments}

\bibliography{mybibfile}

\end{document}